\documentclass[aps,pre,reprint,groupedaddress]{revtex4-1}
\usepackage{graphicx}
\usepackage{float}

\begin{document}


\title{Temperature dependent divergence of thermal conductivity in momentum conserving 1D lattice with asymmetric potential}


\author{Archana G R and Debashis Barik}
\email[]{dbariksc@uohyd.ac.in}
\affiliation{School of Chemistry, University of Hyderabad, Gachibowli, 500046, Hyderabad, India}


\date{\today}

\begin{abstract}
In this study we used nonequilibrium simulation method to investigate the temperature dependent divergence of thermal conductivity in one dimensional momentum conserving system with asymmetric double well nearest-neighbor interaction potential. We show that the value of divergence exponent ($\alpha$) in the power law divergence of thermal conductivity depends on the temperature of the system. At low and high temperatures $\alpha$ reaches close to $\sim0.5$ and $\sim0.33$ respectively. Whereas in the intermediate temperature the divergence of thermal conductivity with the chain length saturates with $\alpha\sim0.07$. Subsequent analysis showed that the predicted value of $\alpha$ in the intermediate temperature may not have reached its thermodynamic limit. Further calculations of local $\alpha$ revealed that its approach towards the thermodynamic limit crucially dependent on the temperature of the system. At low and high temperatures local $\alpha$ reaches its thermodynamic limits in shorter chain lengths. On the contrary in case of intermediate temperature it's progress towards the asymptotic limit is nonmonotonous. 
\end{abstract}


\maketitle

\section{Introduction}\label{intro}

Understanding the heat conduction in finite dimensional systems has been a long standing and much debated problem for some time. Particularly a major focus has been establishing Fourier's law of heat conduction for low dimensional systems. In Fourier's law the heat flux ($J$) becomes proportional to the temperature gradient ($\bigtriangledown T$) with thermal conductivity ($\kappa$) as the proportionality constant, $J=-\kappa \bigtriangledown T$. In pursuit of finding out the microscopic basis of macroscopic law of heat conduction, lattice models of one-dimensional (1D) chain of particles connected by nonlinear interaction potentials have been investigated\cite{Lepri2003a,Dhar2008}. These investigations concluded that in 1D momentum conserving nonlinear lattices the heat flux follows a power law scaling with the system size ($N$), $J \sim N^{\alpha-1}$\cite{Lepri1997}, which is a deviation from Fourier's law that dictates $J \sim N^{-1}$. Therefore in the thermodynamic limit of large $N$, the thermal conductivity diverges with the system size and it scales as $\kappa \sim N^\alpha$. Both theoretical calculations \cite{Lepri1998,Prosen2000,Narayan2002,Lepri2003a,Wang2004a,Lee-Dadswell2005,Li2006a,Delfini2006,Mai2006,VanBeijeren2012,Mendl2013,Spohn2014,Lee-Dadswell2015} and numerical simulations \cite{Lepri1997,Lepri1998,Hu2000,Wang2004a,Li2005,Mai2007,Dhar2008,Roy2012a,Li2015} of lattice models with various variants of Fermi-Pasta-Ulam (FPU) interaction potential predicted anomalous nature of $\kappa$ in 1D momentum conserving systems. Specific values of $\alpha$ varied from one calculation to other\cite{Dhar2008}. However in all cases the exponent lies in the range of $0\le\alpha\le1$. The specific value of $\alpha$ was found to depend on the nature of nonlinearity in the interaction potential. However generally three different value $\alpha=2/5$\cite{Lepri1998,Lepri2003}, $\alpha=1/3$ \cite{Narayan2002,Wang2004a,Mai2006,Mai2007,Roy2012a} and $\alpha=1/2$ \cite{Lee-Dadswell2005,Delfini2006,VanBeijeren2012,Mendl2013} have been obtained in different calculations. Further momentum conserving chains with asymmetric potentials that allow bond dissociation (e.g. Lennard-Jones or Morse potential) were predicted to show convergent thermal conductivity\cite{Savin2014,Sato2016}. In    

However there has been report of finite thermal conductivity in momentum conserving systems with asymmetric interaction potential \cite{Zhong2012a}. This result was in contradiction with theoretical and numerical results in 1D momentum conserving systems that was predicted to have divergence conductivity. The prediction of normal thermal conductivity for asymmetric lattice in 1D at low temperature was debated and possibility of strong low-temperature finite-size effects were discussed by Das \textit{et al.} \cite{Das2014}. Further using equilibrium Green-Kubo method of heat current auto correlation Wang \textit{et al.} \cite{Wang2013} predicted length scale ($N$) dependence of the divergence exponent ($\alpha$) for single-well asymmetric FPU-$\alpha\beta$ and LWAII \cite{Zhong2012a} 1D lattices. They concluded that $\alpha$ reaches its asymptotic thermodynamic limit at a much longer chain length in asymmetric LWAII lattice as compared to asymmetric FPU-$\alpha\beta$ lattice. Therefore at shorter length scale $\alpha$ becomes a function of $N$. However $N$ dependence of $\alpha$ may further be affected by parameters in the potential and more importantly the average temperature of heat bath. Temperature dependent divergence was reported in symmetric double well (DW) potential where $\kappa$ diverged with $\alpha=0.33$ at high temperature and at low temperature it showed weak divergence\cite{Roy2012a}. Similar conclusion was also made for FPU-$\alpha\beta$ lattice with $\alpha=0.4$ at high temperature and weak divergence at low temperature\cite{Savin2014}. Thus temperature dependent divergence in asymmetric momentum conserving lattice needed further systematic investigations. 

In this paper we addressed the problem of temperature dependent divergence of thermal conductivity in 1D nonlinear lattice with asymmetric nearest-neighbor interaction potential. We used asymmetric DW potential for nearest-neighbor interaction among the particles in the lattice. We varied the heat bath temperatures and using nonequilibrium simulation method we calculated $\kappa$ for the lattice of various sizes with different degree of asymmetry in the interaction potential to investigate the temperature dependence of divergence of $\kappa$.   

\section{Model and Results}
We have considered one dimensional lattice model with nearest-neighbor interaction potential. The classical Hamiltonian for the model can be represented as

\begin{equation}
H=\sum_{i=1}^{N}\frac{p_i^2}{2m}+\sum_{i=1}^{N-1}V(x_i-x_{i-1}),
\label{eq1}
\end{equation}

where $x_i$ and $p_i$ are the displacement from equilibrium position and momentum of the $\it{i}$-th particle, respectively. The mass and the total number of particles on the chain are given by $m$ and $N$, respectively. The nearest-neighbor interaction potential between the particle $i$ and $i-1$ is given by $V(x_i-x_{i-1})$. As we have not considered any external potential, our lattice model becomes a momentum conserving chain. We fixed $m=1$ for our calculations throughout. We have chosen asymmetric double well nearest-neighbor interaction potential of the form given by

\begin{equation}
V(x)=-\frac{1}{2}k_2x^2+\frac{1}{3}k_3x^3+\frac{1}{4}k_4x^4,
\label{eq2}
\end{equation}
 
where $k_2$, $k_3$ and $k_4$ are three positive constants. This potential  belongs to the general class of FPU-$\alpha\beta$ potential and due to the cubic nonlinearity the potential becomes asymmetric ($V(x)\neq V(-x)$). In Fig.1 we present the asymmetric nature of double well potential for two different values of cubic nonlinear parameter $k_3$ that determines the degree of asymmetry of the potential. With increase in the value of $k_3$ the asymmetric nature of the potential increases.

In order to study the thermal conduction through the nonlinear chain using nonequilibrium simulation method, both the ends of lattice are connected to Langevin heat baths having different temperatures. The equation of motion of the $i^{th}$ particle in the chain is given by

\begin{eqnarray}
\ddot{x_i} & = & k_2(2x_i-x_{i+1}-x_{i-1})-k_3[(x_i-x_{i-1})^2\nonumber \\
& - &(x_{i+1}-x_i)^2]- k_4[(x_i-x_{i-1})^3-(x_{i+1}-x_i)^3] \nonumber \\
& - &\gamma_i\dot{x_i}+\eta_i ,
\label{eq3}
\end{eqnarray}

where the fluctuation ($\eta_i$) and dissipation ($\gamma_i$) terms are defined as $\eta_i=\eta_L\delta_{i,1}+\eta_R\delta_{i,N}$ and $\gamma_i=\gamma(\delta_{i,1}+\delta_{i,N})$, respectively. The heat baths are characterized by the fluctuation-dissipation relations followed by the two Markovian heat baths, $\langle \eta_L(t)\eta_L(t^\prime)\rangle=2\gamma k_BT_L\delta(t-t^\prime)$ and $\langle \eta_R(t)\eta_R(t^\prime)\rangle=2\gamma k_BT_R\delta(t-t^\prime)$. The $\gamma$, $k_B$, $T_L$ and $T_R$ are the dissipation constant, Boltzmann constant, temperatures of left and right heat baths, respectively. The values of Boltzmann constant ($k_B$) and the dissipation constant ($\gamma$) were chosen to be unity throughout. We varied the left and right bath temperatures ($T_L$ and $T_R$) to investigate the effect of temperature on the divergence behavior of thermal conductivity. In this context we defined two relevant quantities: the temperature difference, $\Delta T=T_L-T_R$ and the average temperature, $T=(T_L+T_R)/2$. 

The instantaneous local heat current between $i^{th}$ and $(i+1)^{th}$ is defined by 

\begin{eqnarray}
j_i=\frac{1}{2}(\dot{x}_i+\dot{x}_{i+1})\frac{\partial H}{\partial x_i} .
\label{eq4}
\end{eqnarray}

Defining the time-averaged local heat current as $J_i=\lim_{t\rightarrow\infty}\frac{1}{t}\int_{0}^{t}j_i(\tau)d\tau$, that reaches a nonequilibrum stationary state across the lattice after long time, the global heat current in the lattice is given by

\begin{eqnarray}
J=\sum_{i=1}^{N-1}\frac{J_i}{N-1} .
\label{eq5}
\end{eqnarray}

The thermal conductivity is related to the steady state global heat current as

\begin{eqnarray}
\kappa=\frac{JN}{\Delta T} .
\label{eq6}
\end{eqnarray}

If the global heat current follows $J\sim N^{-1}$ scaling then the thermal conductivity becomes convergent in the thermodynamic limit (large $N$). However in 1D momentum conserving systems $J$ has been predicted to scale as $J\sim N^{\alpha-1}$. Thus $\kappa$ becomes divergent with a power law scaling relation as $\kappa \sim N^\alpha$.  

\begin{figure}
	\centering
	\includegraphics[trim = 55mm 60mm 60mm 45mm,clip,width=1\linewidth]{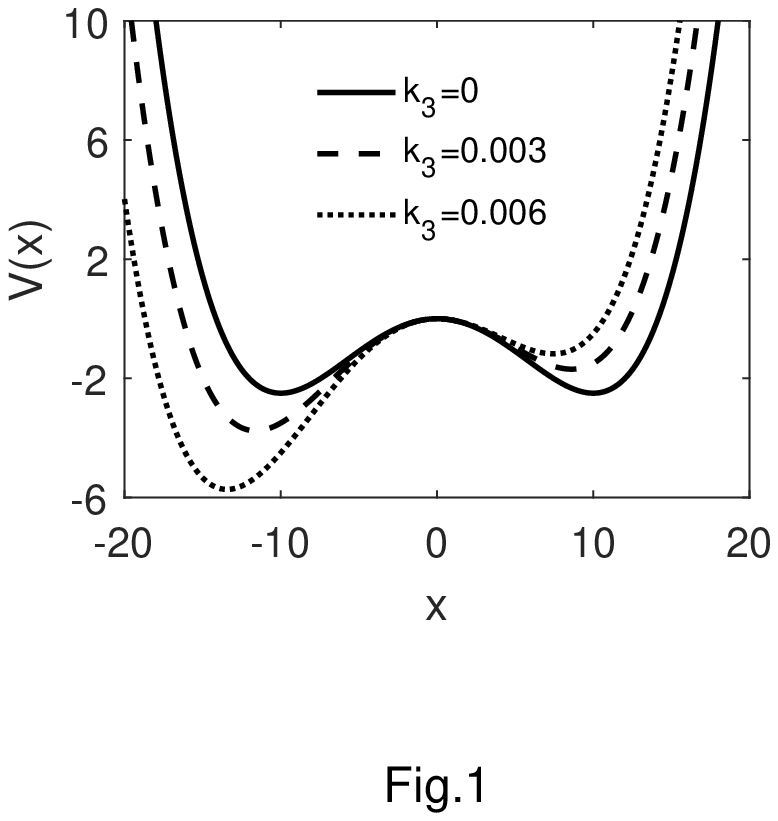}
	\caption{Schematic representation of the double-well (DW) nearest-neighbor interaction potential, $V(x)$, with different values of asymmetric parameter $k_3$. $k_2=0.1$ and $k_4=0.002$.}
	\label{fig:fig1}
\end{figure}

In order to numerically integrate the dynamical equations (\ref{eq3}), we used 4-th order Runge-Kutta method. We chose to use 4-th order Runge-Kutta method to achieve higher accuracy in our calculations although it leads to considerable increase in simulation time. We typically ran $2-5\times10^{10}$ iterations in numerical integrations with an integration step length of $0.01$. In the DW potential (\ref{eq2}) we fixed $k_2=0.1$, $k_4=0.002$ and varied $k_3$ ($0.003$ or $0.006$) in order to explored the effect of asymmetry on the nature of divergence in $\kappa$. Further to investigate the temperature dependence of divergence we chose various combination of $T$ and $\Delta T$ ($T, \Delta T$). We used fixed boundary condition (BC), $x_0=x_{N+1}=0$, in our calculations. Previous works highlighted the importance of boundary conditions in the heat conduction in lattice models\cite{Das2014,Cividini2017}. 

In Fig.2a we show the divergence of thermal conductivity for the asymmetric potential ($k_3=0.003$) with varying average heat bath temperatures keeping the $\Delta T$ fixed. The chosen ($T$, $\Delta T$) pairs were ($9.5, 1.0$); ($4.5, 1.0$); ($1.5, 1.0$). For these three different values of $T$ the system exhibits power-law divergence of thermal conductivity with $\alpha$ ranging between $0.31-0.35$. These $\alpha$ values are similar to the predicted $\alpha=1/3$ by renormalization group theory, mode coupling theory and many numerical simulations\cite{Narayan2002,Wang2004a,Mai2006,Mai2007,Roy2012a,Delfini2006}. We found similar divergence of $\kappa$ for the same system with higher asymmetry ($k_3=0.006$) in the interaction potential (Fig.2b). One important aspect of these divergence behavior is that the average temperature of the system is large. Therefore at the high temperature limit the asymmetric-DW-momentum-conserving system behaves similar to the symmetric-FPU-$\alpha\beta$-momentum-conserving system. 

\begin{figure}
	\centering
	\includegraphics[trim = 45mm 50mm 50mm 40mm,clip,width=1\linewidth]{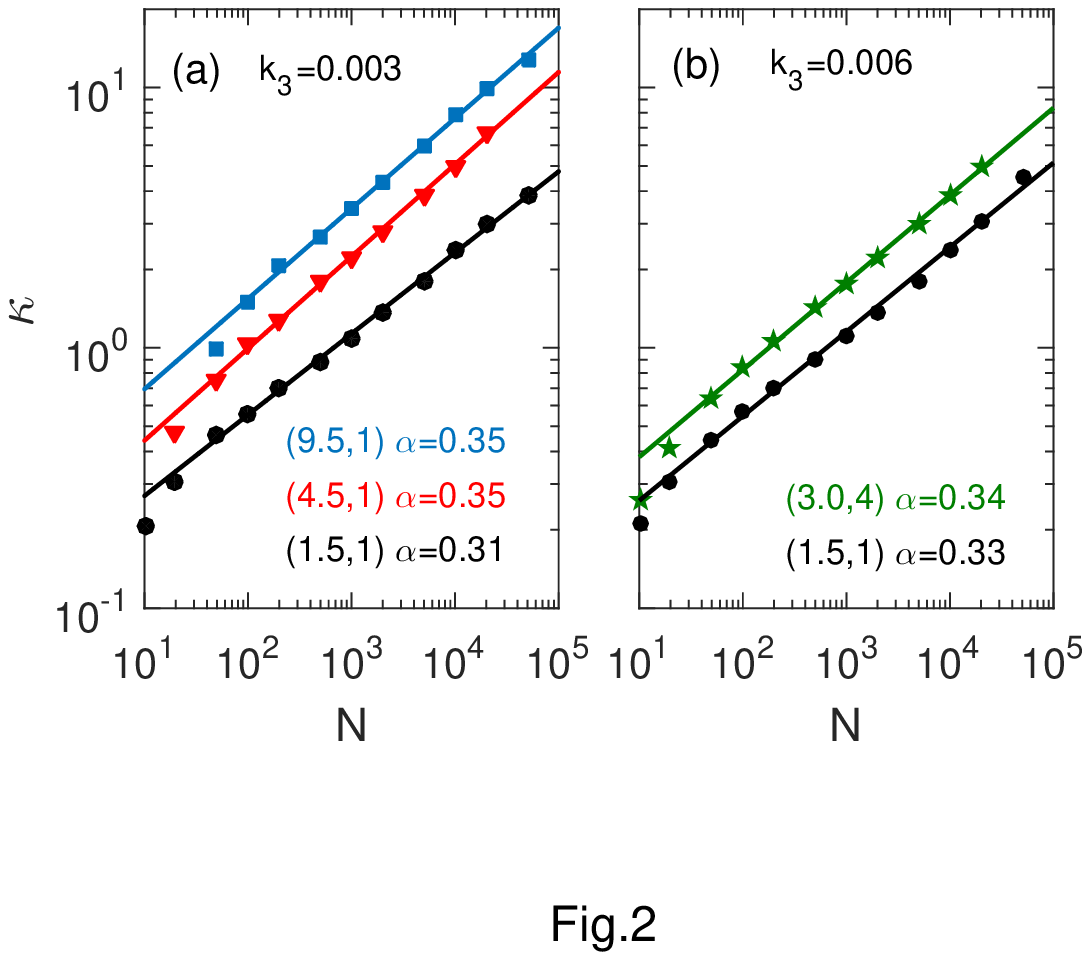}
	\caption{Divergence of thermal conductivity, $\kappa$, as a function of chain length, $N$. Different colored symbols represent simulations with different average bath temperatures with fixed temperature difference, ($T$, $\Delta T$); circle: ($1.5,1$), triangle: ($4.5,1$), square: ($9.5,1$) and star: ($3.0,4$). Solid lines are from power law fitting ($\kappa\sim N^\alpha$). The values of $\alpha$ are indicated inside the plots for (a) $k3=0.003$ and (b) $k3=0.006$.}
	\label{fig:fig2}
\end{figure}

\begin{figure}
	\centering
	\includegraphics[trim = 60mm 60mm 60mm 40mm,clip,width=1\linewidth]{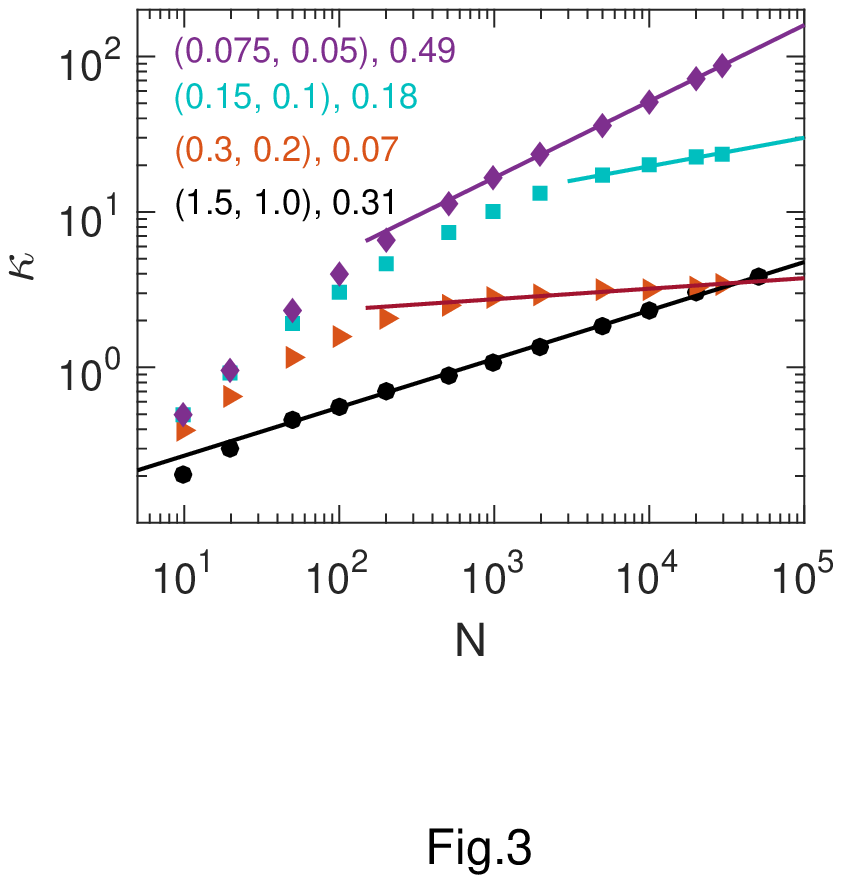}
	\caption{Divergence of $\kappa$ as a function of $N$ for different average $T$ and $\Delta T$;  circle: ($1.5,1$), triangle: ($0.3,0.2$), square: ($0.15,0.1$) and diamond: ($0.075,0.05$). The asymmetric parameter $k_3$ was $0.003$. The $\alpha$ values are indicated inside the plot.}
	\label{fig:fig3}
\end{figure}

\begin{figure}
	\centering
	\includegraphics[trim = 53mm 55mm 65mm 45mm,clip,width=1\linewidth]{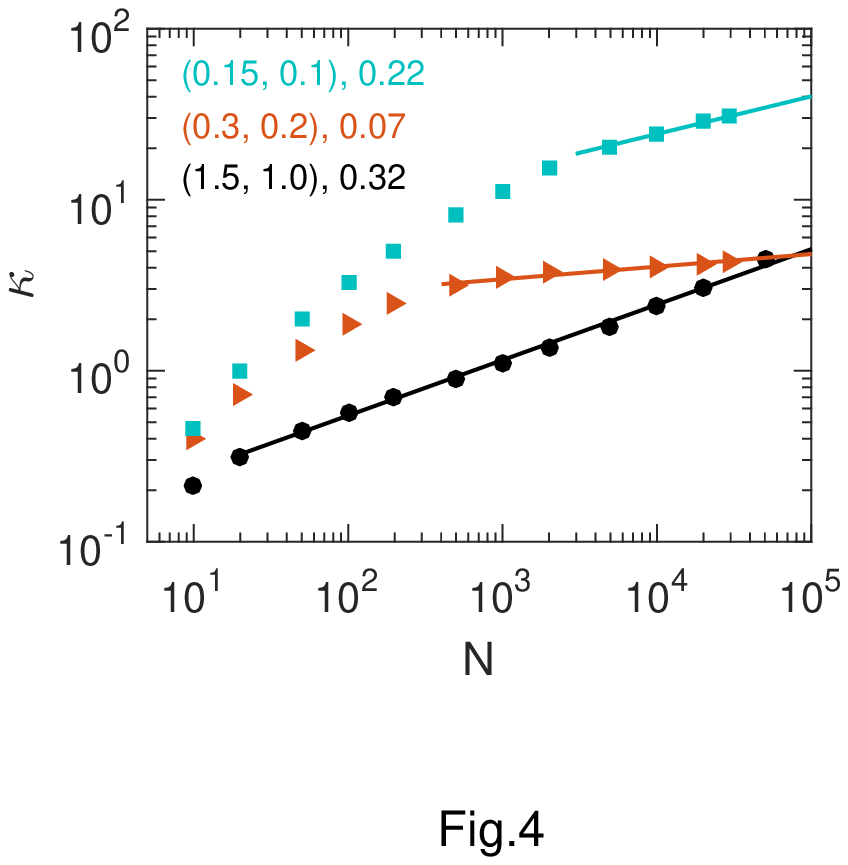}
	\caption{Divergence of $\kappa$ with $N$ at different temperatures with asymmetric parameter $k_3=0.006$. ($T$, $\Delta T$) pairs are: circle-($1.5,1$), triangle-($0.3,0.2$), square- ($0.15,0.1$).}
	\label{fig:fig4}
\end{figure}

We next investigated the divergence behavior of thermal conductivity for a range of average temperature values in the intermediate to low $T$ limits again by varying the heat bath temperatures $T_L$ and $T_R$. Particularly we aimed to determine the nature of divergence in the intermediate and low temperature regimes. In Fig.3 we show the divergence of $\kappa$ in different average temperatures of the system with $k_3=0.003$. Fig.3 indicates that the qualitative nature of divergence changes depending on the average temperature of the heat bath. The conductivity diverged sharply with $\alpha=0.49$ at very low temperature ($T=0.075$). With increase of temperature ($T=0.15$) the divergence becomes shallow with $\alpha=0.18$. Further increase of temperature ($T=0.3$) the thermal conductivity appears to saturate with $N$ with $\alpha=0.07$. At high temperature ($T=1.5$) $\kappa$ shows its usual divergence behavior with $\alpha=0.31$. The striking feature of the temperature dependent thermal conductivity here is that two different types of scaling behaviors of $\kappa$ at very low ($\alpha=0.49$) and very high ($\alpha=0.31$) temperatures. Further the very weak divergence of thermal conductivity (or saturation of $\kappa$ with $N$) in the intermediate $T$ poses a possibility of validity of Fourier's law. In addition, consistent with previous observations \cite{Das2014,Savin2014} we also find the weak divergence with the fixed BC even though fixed BC does not allow thermal expansion that was known to provide additional avenue of phonon scatterings leading to normal conductivity. Similar saturation of $\kappa$ with $N$ was first reported in asymmetric lattice by Zhong \textit{et al.}\cite{Zhong2012a} and it was proposed by them that the system follows Fourier's law. However in asymmetric FPU-$\alpha\beta$ potential whether the saturation was indeed due to asymmetric nature of interaction potential or not was discussed later \cite{Das2014,Wang2013}. Repeating calculations with higher asymmetry of the potential ($k_3=0.006$) also resulted similar observations as in $k_3=0.003$ (Fig.4). Thus our simulation results indicate that the values of $\alpha$ depends on the temperature of the system in asymmetric DW interaction potential.

\begin{figure}
	\centering
	\includegraphics[trim = 55mm 65mm 60mm 40mm,clip,width=1\linewidth]{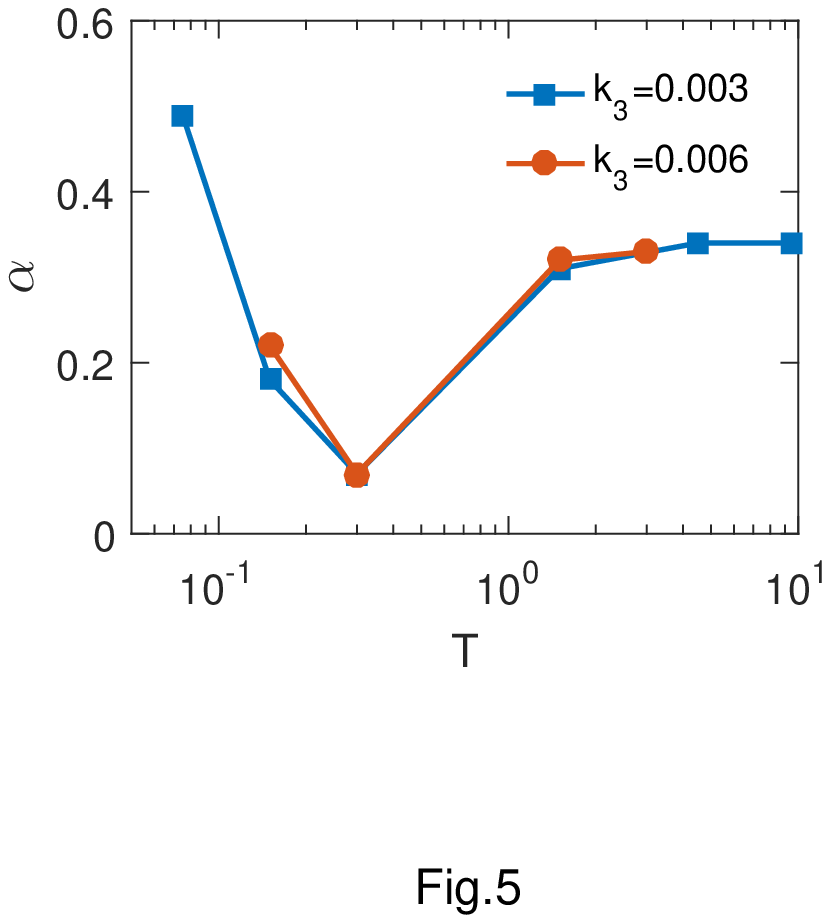}
	\caption{Temperature dependence of $\alpha$ for two different values of asymmetric parameters $k_3$.}
	\label{fig:fig5}
\end{figure}

To determine the temperature dependence of $\alpha$, we plotted the $\alpha$ as a function of $T$ for two different values of asymmetric parameter $k_3$ (Fig.5). With increase in $T$, $\alpha$ decreases sharply and passing through a minimum it increases to saturate with $\alpha=0.35$ at high $T$. The weakest divergence of $\kappa$ occurs at the intermediate $T$ both for low and high asymmetries of the potential. This type of turnover behavior has been reported in the recent past \cite{Xiong2014,Xiong2016} in case of 1D anharmonic chain. The comparison of $\alpha$ vs. $T$ for low and high $k_3$ indicates that the divergence behaviors of thermal conductivity for different asymmetry values are identical. If the saturation of $\kappa$ in this system was an asymmetry induced effect then there must have been a shift in $\alpha$ vs. $T$ plots for the two different values of $k_3$. However the two curves overlap with each other. Further for the same reason, expectedly higher asymmetry would have resulted saturation of $\kappa$ at lower $N$ as compared to lower asymmetry \cite{Das2014}. The comparison of $\kappa$ vs. $N$ profiles for higher and lower asymmetry at different $T$ (Fig.6) do not indicate any such asymmetry induced early saturation of $\kappa$. These results and analyses point out that the saturation of $\kappa$ may be a finite length effect occurs only at intermediate $T$. However our results does point out that the nature of divergence is indeed temperature dependent.

\begin{figure}
	\centering
	\includegraphics[trim = 56mm 65mm 60mm 40mm,clip,width=1\linewidth]{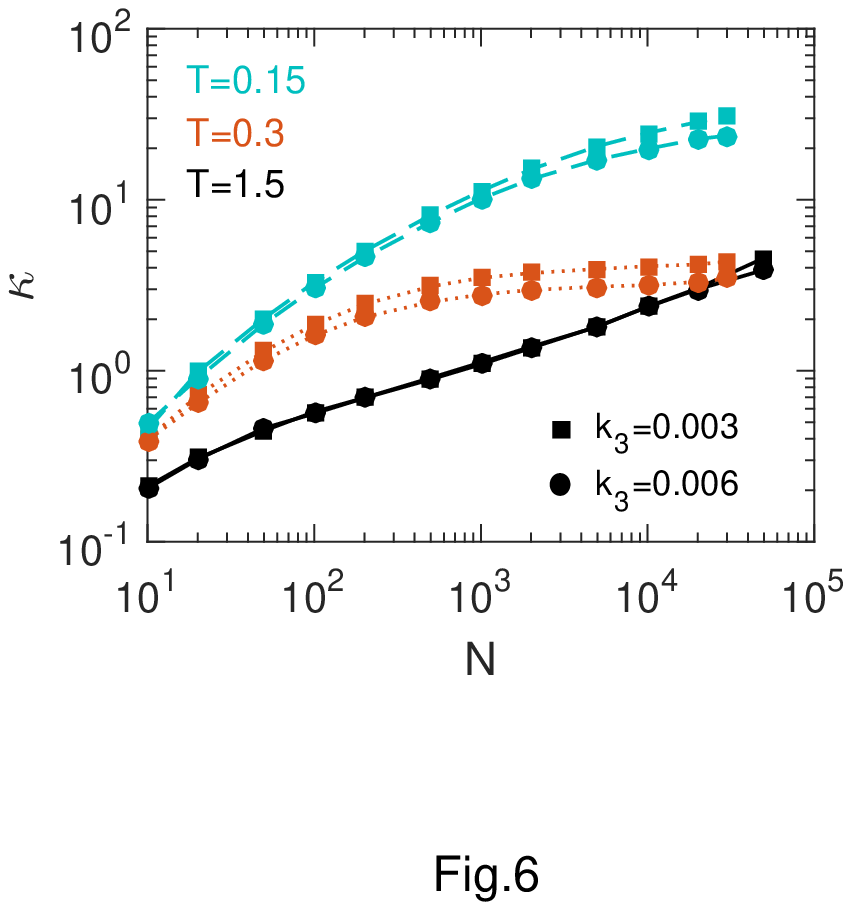}
	\caption{Comparison of divergence of $\kappa$ with $N$ for different values of asymmetric parameter $k_3$ at various average bath temperatures, $T$. Solid line: $T=1.5$, dotted line: $T=0.3$ and dashed line: $T=0.15$.}
	\label{fig:fig6}
\end{figure}

\begin{figure}
	\centering
	\includegraphics[trim = 53mm 50mm 65mm 50mm,clip,width=1\linewidth]{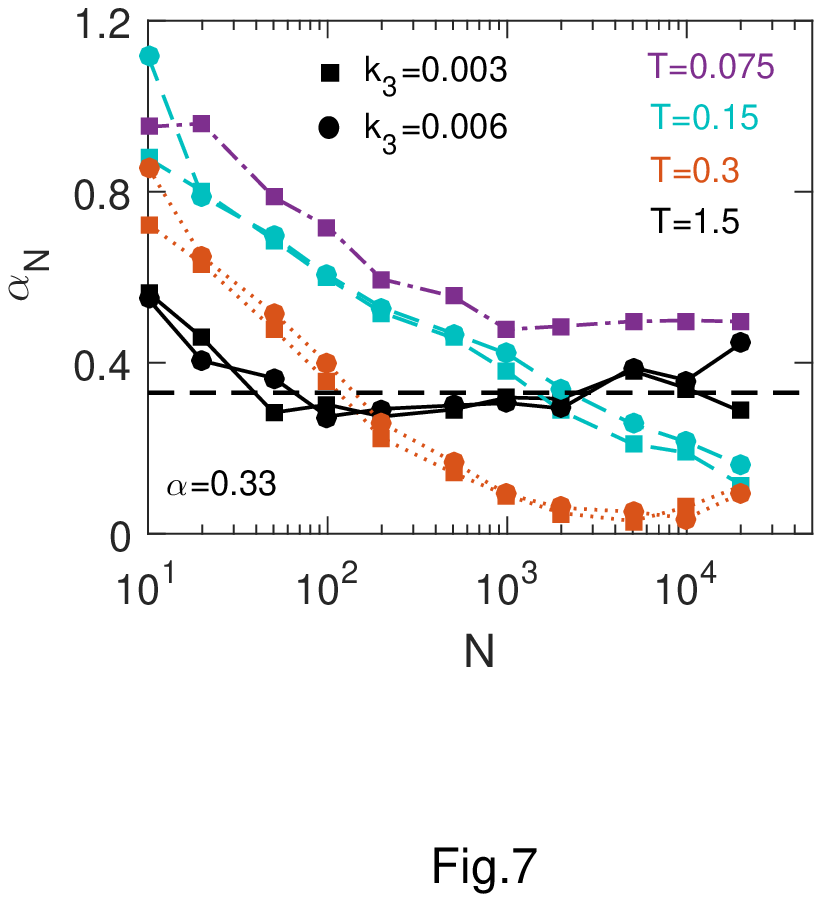}
	\caption{Plot of local divergence coefficient $\alpha_N$ with $N$ for different values of asymmetric parameter $k_3$ and at different average bath temperature, $T$. The $\alpha_N$ was estimated by calculating the local slope of $\kappa$ vs. $N$ plots given in Fig.3 and Fig.4. Solid line: $T=1.5$, dotted line: $T=0.3$, dashed line: $T=0.15$ and dashed-dot line: $T=0.075$. The horizontal dashed line represents $\alpha=0.33$.}
	\label{fig:fig7}
\end{figure}

In order to determine the finite-size effect on $\alpha$ we calculated the local divergence coefficient, $\alpha_{N}$, by determining the local slope in $\kappa$ vs. $N$ line. In Fig.7 we present the $\alpha_N$ as a function of $N$ estimated at various $T$ for two different values asymmetry parameter. At high $T$ ($T=1.5$) the well-known thermodynamic limit of $0.33$ is achieved at the shorter length of the chain and $\alpha_N$ settles nearly with that value for large range of $N$.  On the other hand, at very low $T$ ($T=0.075$), $\alpha_N$ appeared to settle at $\alpha\sim0.5$ value, indicating the different scaling behavior of the system depending on the temperature of the system. However at the intermediate $T$ ($T=0.3$) with increasing $N$, $\alpha_N$ decreases below the thermodynamic limit ($\alpha=0.33$, dashed line in Fig.7) and passing through a minimum it shows an increasing trend for both the values of asymmetry parameters. Similar trend was seen for $T=0.15$ although without the minimum as probably the minimum is located at larger $N$. As at these two temperatures the local $\alpha$ does not settle to a particular value, it may be concluded that the value of $\alpha$ at the intermediate $T$ are not from the thermodynamic limit of the system. 

\section{Conclusion}  

Understanding the divergent nature of thermal conductivity in low dimensional systems has been a long standing problem. A large number of theoretical and numerical calculations on 1D momentum conserving systems concluded power-law divergence of thermal conductivity with the length of lattice \cite{Dhar2008,Lepri2003a}. In this study we used nonequilibrium simulation method to show that the divergent nature of $\kappa$ in 1D asymmetric lattice depends on the temperature of the heat baths. In the thermodynamic limit, the system exhibits $\alpha\sim0.5$ and $\alpha\sim0.33$ at low and high $T$ respectively. Therefore our calculations points out two different scaling behavior of the same system depending on the temperature of the system. At low $T$ our predicted value of $\alpha$ ($\sim0.5$) becomes same as given previously by mode coupling theory \cite{Lee-Dadswell2005,Delfini2006,VanBeijeren2012,Mendl2013}. Whereas at high $T$ our calculation leads to an $\alpha$ ($\sim0.33$) as predicted previously by renormalization group analysis \cite{Narayan2002,Mai2006}. Further at the intermediate $T$, $\kappa$ appears to saturate against $N$ with very small value ($\alpha=0.07$). Similar weak divergence of $\kappa$ has been reported before in case of 1D asymmetric momentum conserving lattice \cite{Zhong2012a} and it was characterized as the validity of Fourier's law by the asymmetric system. However latter it was determined that the behavior may not be associated with the \textit{true} thermodynamic limit of the system \cite{Das2014,Wang2013}. In order to probe the weak divergence of $\kappa$ in the intermediate $T$,  we calculated local divergence exponent, $\alpha_N$, and showed that in the intermediate $T$, $\alpha_N$ does not saturate to a fixed value in the length scale of our simulations. $\alpha_N$ decreases with $N$ and passing a minima it showed a tendency to increase again with $N$. Had the system reached its thermodynamic limit there should not have any further increase in the local $\alpha$. On the contrary in the low and high temperatures $\alpha_N$ decreases with $N$ and saturate to its respective thermodynamic limits independent of extent of asymmetry in the interaction potential. Therefore our calculations indicate that the approach to the thermodynamic limit of $\alpha$ is indeed temperature dependent in case of asymmetric interaction potential.

\begin{acknowledgments}
This work is partially supported by Major Research Project scheme of University Grants Commission, India (MRP-MAJOR-CHEM-2013-43860).
\end{acknowledgments}


%

\end{document}